# Effect of hydrostatic pressure on irreversible thermal transformations in a polymer glass at low temperatures

J. Takahashi,[1,2,*] A. Suisalu,[2] An. Kuznetsov,[2] A. Laisaar,[2]
V. Hizhnyakov,[1,2] and J. Kikas[1,†]

[1]*Department of Physics, University of Tartu, Tähe Street 4, 51010 Tartu, Estonia*
[2]*Institute of Physics, University of Tartu, Riia Street 142, 51014 Tartu, Estonia*
(Dated: 19 January 2004;  submitted to *Phys. Rev. B*)

Irreversible broadening of spectral holes in chlorin-doped polystyrene glass was studied for the first time in the temperature cycling experiments *under high pressure* (by raising the temperature from 5 K to various magnitudes up to 18 K and turning back to 5 K at several fixed pressures between 0 and 5 kbar). At all pressures the increment in the hole width observed after completing a temperature cycle exhibits a slightly superlinear ($\propto T^{3/2}$) dependence on the cycling temperature. The magnitude of this increment is essentially reduced under high pressure (e.g., at 4.9 kbar it makes up less than 2/3 of its initial value obtained at ambient pressure). The residual broadening of holes is interpreted as a result of irreversible thermally induced spectral diffusion arising from interaction of the electronic transition in a dopant molecule with two-level systems (TLSs) which perform thermally activated overbarrier jumps between two possible states of a TLS. The pressure effects are treated theoretically within the scope of the soft anharmonic potential model with asymmetric distribution of the cubic anharmonicity parameter. It is shown that more abundant are TLSs with such double-well potentials where the minimum, corresponding to a larger glass volume, is placed at a higher energy than the minimum, corresponding to a smaller volume. In this case, applied pressure reduces the number of almost symmetric TLSs (having very small energy difference between the two minima), which provide the largest contribution to the residual broadening of spectral holes after a temperature cycle. Our earlier results on isothermal hole burning at various fixed pressures [V. Hizhnyakov *et al*., Phys. Rev. B **62**, 11296 (2000)] also qualitatively fit into this picture.

PACS numbers: 71.55.Jv,  62.50.+p,  71.23.An,  78.40.Me

---

[*]Present address: Japan Science and Technology Corporation (CREST), 758-65 Bibi, Chitose, Hokkaido 066-8655, Japan
[†] Electronic address: jaakk@physic.ut.ee

# I. INTRODUCTION

A major progress in the physics of glasses is related to the discovery in the early 1970s of low-temperature anomalies in amorphous solids.[1] Extra heat capacity, reduced thermal conductivity and several other phenomena, such as enhanced ultrasound and microwave attenuation, acoustic echoes etc., distinguish glasses from crystalline materials. The tunneling model, based on an idea of two-level systems (TLSs),[2,3] was proposed in order to explain these anomalies. This model was later extended to the soft anharmonic potential (SAP) model[4-6] which allows a unified treatment of tunneling TLSs as well as of other specific "glassy" excitations – soft localized (vibrational) modes (SLMs). Both these low-energy excitations, being characteristic for all glasses, manifest themselves in various physical effects. Their coupling to the optical transitions in probe molecules doped into a glass is revealed in low-temperature ambient-pressure spectral hole burning[7] and photon echo[8] experiments as peculiar slightly superlinear temperature dependences of hole widths and echo dephasing rates.

Our group was the first in extending the spectral hole burning (SHB) studies into the high-pressure domain above 1 kbar[9] (see also our succeeding original papers[10-13] and review articles[14-17] on the same subject). Pressure allows to alter various interactions in a sample under study by changing its interatomic distances directly and smoothly, as opposed to stepwise variation of this important parameter by selecting a series of analogous systems having different chemical composition.[18] It should be stressed that in all our experiments only moderately high hydrostatic pressures (up to ~8 kbar) were used. Utilizing such not very high pressures makes it possible to investigate in detail a pure linear response of an organic glass to the external pressure. This offers certain advantages over experiments with diamond anvil cells where one deals with much higher pressures of tens and hundreds kilobars. At such strong compressions various nonlinear pressure effects come into play and it makes interpretation of obtained results much more difficult.

The aim of this paper is to study the effect of hydrostatic pressure on *irreversible* processes in a glass at liquid helium temperatures. A sensitive method to trace irreversible thermal transformations in solids is offered by temperature cycling of spectral holes.[19,20] Roughly speaking, an irreversible residual broadening of a spectral hole resulting from such a temperature cycle is related to the number of transformed TLSs, i.e. to those TLSs which have changed their "well of location" in this cycle (recall that all TLSs are described by double-well potentials in the SAP model). A pronounced suppression of the cycling-induced hole broadening, observed in this study even at relatively low pressures of some kilobars, indicates that the number of such TLSs is remarkably reduced by the applied pressure. The suppression is proportional to pressure $P$ in the low $P$ limit. This linear pressure effect corroborates the existence of linear coupling of TLSs to the pressure-induced strain. We will demonstrate that the effect can be considered within the general theoretical framework of the SAP model, which was previously applied to our isothermal SHB experiments under pressure.[11-13] However, some modification of this model is necessary, as will be shown below.

## II. EXPERIMENT

The system under study was glassy polystyrene (PS) doped with free-base chlorin (Chl), a synthetic organic dye from the class of porphyrins. A molecule of chlorin (7,8-dihydroporphin, $H_2 - C_{20}N_4H_{14}$) is a very convenient probe for SHB studies owing to existence of an effective intramolecular phototransformation process (turning of the central proton pair by $90°$ as a result of absorption of a photon by the molecule). The content of dopant in the sample was estimated to be about $1.2 \times 10^{-7}$ moles of Chl per gram of PS or $1.3 \times 10^{-4}$ moles per $dm^3$ ($1.3 \times 10^{-4}$ M). The sample was prepared by evaporation of a low-molecular-weight solvent from a solution of polymer and dye followed by a prolonged drying at normal conditions. The obtained sheet of polystyrene was about 0.2 mm thick and had a greenish hue. From this sheet the specimen of ~5×5 mm size was cut. For hole burning we used the longest-wavelength $S_1 \leftarrow S_0$ absorption band, $Q_x$ (0 – 0) band,[21] of chlorin molecules peaked at ~636 nm under normal pressure at liquid helium temperature. The optical density of our specimen in the maximum of this band was about 1 (see Fig. 1) as was determined in separate experiments (not described here in detail) by measuring the transmission spectra at 50 bar and 3.7 kbar with the aid of a low-resolution tunable dye laser (Coherent CR-490) having linewidth ~0.5 $cm^{-1}$.

All the hole burning and measuring procedures were carried out with a very high resolution Coherent CR-699-29 single-frequency tunable ring dye laser of ≤2 MHz linewidth by using DCM dye. Irrespective of the magnitude of pressure applied in a particular experiment, holes were always burned at around 15680 $cm^{-1}$, i.e. slightly to the red from the peak of the $Q_x$ band at normal pressure. In fact, the absorption band shifts under applied pressure to the red at a rate $d\mathbf{n}/dP$ of about $-7.7$ $cm^{-1}$/kbar for the band peak as can be seen from Fig. 1. (Almost the same value, namely $-7.8$ $cm^{-1}$/kbar, was also obtained from a separate pressure-tuning experiment at low pressures 20 – 60 bar of gaseous helium (see below), by measuring the pressure shift of a hole burned at the band peak as described in Ref. 10.) However, the shift of this band even at the highest pressure used (5 kbar) is small as compared to the total bandwidth (~40 $cm^{-1}$ *versus* ~200 $cm^{-1}$, see Fig. 1) and can hardly contribute to the effects observed (via the wavelength dependence of the residual broadening of holes in temperature cycling experiments). Note also that the width of spectral holes in isothermal experiments is practically independent of the burning wavelength all over the absorption band as was ascertained by us earlier.[9]

The holes under study were recorded in the transmission mode. Because a burned hole becomes more and more shallow after each temperature cycle, deep holes with the optical density decrements of 0.2 to 0.3 were burned. The hole profiles were fitted to the Lorentzian function and the full width at half maximum (FWHM) of the holes was obtained by using the fitting parameters. Since our holes were burned rather deep, their shapes somewhat deviated from Lorentzian curves. However, it was estimated that this deviation did not essentially affect the obtained results.

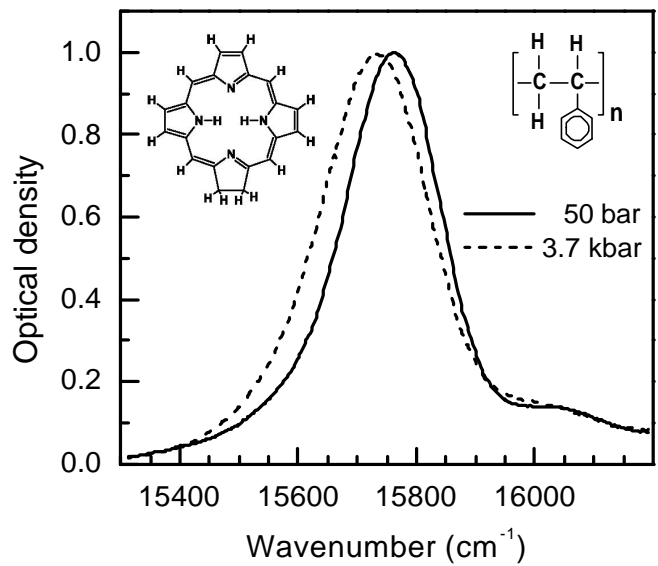

FIG. 1. Absorption spectra of chlorin probe molecules in a glassy polystyrene at $T = 4.8$ K under two different pressures: 50 bar and 3.7 kbar. The insets show structural formulas of a chlorin molecule (left) and of polystyrene (right).

A high hydrostatic pressure was applied to the sample by using a special complex of equipment,[9,10] consisting of a 15-kbar gas compressor, a high-pressure optical cell with three sapphire windows and a temperature-controlled liquid-helium immersion-flow cryostat. The sample under study was mounted free of strain in the pressure cell and after that the cell was installed inside the cryostat. Then compressed helium gas, used as the pressure-transmitting medium, was fed from the compressor to the pressure cell via a flexible capillary tube. Note that this operation was done at room temperature and hence well below the glass transition temperature $T_g$ of polystyrene, which is approximately 90–100°C at atmospheric pressure, depending on the sample prehistory and some other factors (see, e.g., Refs. 22,23). The pressure of gaseous helium in the pressure cell was measured with a manganin resistance gauge located inside the compressor and thus held at room temperature. Pressures up to 5 kbar were employed in the present investigation.

The next step was cooling a rather massive pressure cell (~4.6 kg; Ø100x80 mm) from room temperature down to about 80 K by using liquid nitrogen. This procedure took ordinarily about 2 hours. After that the cryostat was loaded with liquid helium, and within 15 to 20 minutes the temperature of the pressure cell reached 5 K, the initial temperature of all our temperature cycling experiments. Note that we did not go down to 4.2 K, the boiling temperature of liquid $^4$He at atmospheric pressure, because in that case the pressure cell would be surrounded by liquid helium and therefore the intensity of light beams passing through the cryostat would be affected by gas bubbles released from boiling helium. Instead, the pressure cell resided during the measurements in gaseous helium evaporated from the helium bath. The total time to cool the cell down to the temperature of experiment was almost 3 hours.

Each our cycling (excursion) temperature within the range 5–18 K as well as the starting temperature 5 K of the pressure cell and the sample inside it was stabilized and kept constant with an accuracy of ±0.1 K. The temperature was measured by a calibrated thermocouple (Cu + 0.1 at.% Fe + 0.02 at.% Li *vs.* copper) attached to the outside of the pressure cell. We assume that the sample inside the cell attained practically the same temperature as the thermocouple in the course of temperature stabilization taking some 3-5 minutes.

During the cooling of the cryostat by means of liquid helium, the compressed helium gas solidified (froze) first of all in the capillary tube connecting the pressure cell with the compressor and only afterwards inside the massive pressure cell body. (Notice that at any temperature above $T_{crit} = 5.2$ K, helium-4 exists, depending on *P,T* values, either in the gaseous or a solid phase but never in the liquid state.) Because the capillary became blocked in the course of cooling, the freezing of helium inside the cell proceeded under constant volume conditions (isochorically) and was accompanied by a drop in pressure during the freezing. An additional pressure loss took place in the process of the further cooling of solidified helium along an isochore. The final pressure of solid helium acting on the sample at a fixed temperature (between 5 and 18 K in our study) could not be measured directly by the manganin gauge. Instead, it was estimated by using the most precise available high-pressure low-temperature *P,V,T* data for $^4$He obtained and tabulated in Ref. 24 (see also an earlier paper[25]). For instance, from Table VI of Ref. 24 one can see that by cooling at a pressure of about 1.9 kbar, isochoric freezing of gaseous helium begins at $T_{mf} = 20.61$ K under $P_{mf} = 1860.3$ bar and terminates at $T_{ms} = 17.57$ K under $P_{ms} = 1449.4$ bar. Here subscripts *f* and *s* mean fluid and solid phase, whereas subscript *m* designates that the above points are located on the melting line of $^4$He (see Fig. 7 of Ref. 24). By further cooling along the isochore with the molar volume $V = 11.50$ cm$^3$/mole, the pressure falls to the value $P_0 = 1365.0$ bar at $T = 0$ K.

It should be noted that in each temperature cycle the pressure inside the cell inevitably varies to some extent in response to the temperature variation (due to the thermal expansion/ contraction of compressed solid helium at constant volume). From the data presented in Tables VI and VII of Ref. 24 one can estimate such pressure changes quite accurately employing interpolation with cubic splines. For example, it can be found that at a base pressure of 2 kbar the temperature elevation from 5 K to 15 or 18 K is accompanied by increase in pressure by 35 or 67 bar, whereas at 4 kbar the respective pressure increment is 19 or 37 bar. Thus, the effect of thermal variation of pressure decreases as the base pressure increases.

The thermally induced changes of spectral holes were measured as follows. First, a hole was burned at the lowest temperature of 5 K and measured. Second, the temperature was raised to a desired cycling (excursion) temperature and kept constant for a few minutes. Then the sample was re-cooled to the initial temperature and the hole was measured again (post-cycle measurement). Thereupon the excursion temperature was raised to the next step and so forth, up to 18 K. Notice that the *same* spectral hole was repeatedly recorded after each new temperature cycle. The duration of individual temperature cycles was different, depending on the excursion temperature. For the longest cycle (5 K → 18 K → 5 K) it amounted to about 20 minutes. The temperature as well as the pressure stability was checked by recording the same hole once more, after waiting for some minutes. In fact, this check should reveal any temporal changes in the spectrum of the sample under the given conditions. Nothing of this kind was observed in our experiments.

Since all our high-pressure experiments were carried out at such *P,T* conditions that helium inside the pressure cell remained solid (even at the highest excursion temperature), the temperature alteration at a constant cell's volume had to involve some changes in pressure, as mentioned above. However, as was checked in the separate pressure-tuning experiments at much lower pressures (<100 bar), where helium inside the pressure cell was in liquid or gaseous state (at 4.2 K liquid $^4$He solidifies under pressure of about 140 bar[24]), the changes in the properties of spectral holes (their spectral position, width, area, etc.) were practically entirely *reversible* with respect to pressure variations in this tuning range (see, e.g., our paper[10]). We assume that the same statement holds for the region of higher pressures of some kilobars as well.

## III. THEORY

### A. The soft anharmonic potential model

In the soft anharmonic potential (SAP) model, both low-energy excitations that are peculiar to glasses — two-level systems (TLSs) and soft localized modes (SLMs) — are described by the following anharmonic oscillator potential represented by a polynomial of fourth order:[4-6,26,27]

$$U(x) = U_0\left[h\left(\frac{x}{a}\right)^2 + x\left(\frac{x}{a}\right)^3 + \left(\frac{x}{a}\right)^4\right]. \quad (1)$$

Here *x* is the generalized configurational coordinate presenting the displacement of atoms involved in TLSs and/or SLMs, $U_0$ is the characteristic binding energy of particles in a glass (~10 eV), and *a* is the length unit on the atomic scale. For the sake of simplicity of further presentation and compactness of formulas, we take *a* equal to 1. Then Eq. (1) reads

$$U(x) = U_0\,(h\,x^2 + x\,x^3 + x^4), \quad (2)$$

where $x$ is now dimensionless coordinate which is expressed in the length units $a = 1$ Å and varies, say, between $-0.5$ and $+0.5$.

The quantities $h$ and $x$ are random dimensionless parameters. Due to the spatial inhomogeneity of the glassy structure they are subject to a statistical distribution described by the distribution function $F(h,x)$. Both $h$ and $x$ may have either positive or negative values. For soft potentials they have values much smaller than unity ($|h|, |x| \ll 1$). It is known that different regions in the ($x,h$) plane correspond to different types of low-energy excitations. Namely, one gets TLSs, which are always described by a *double-well* potential, in the following cases: (i) when $h > 0$ and it satisfies the condition $h < 9x^2/32$ for both positive and negative $x$ values, and (ii) when $h < 0$ and $x$ has any positive or negative value. On the other hand, SLMs, which are represented by a *single-well* potential, exist only when $h > 0$ and the reverse inequality $h > 9x^2/32$ is satisfied for all positive and negative $x$ values.

Behavior of SAPs, both SLMs and TLSs, in a glass under high hydrostatic pressure has already been considered in our earlier papers[11-13] (see also reviews[14-17]) in connection with the investigation of the effect of pressure on the homogeneous width of zero-phonon lines at liquid helium temperatures as determined by the SHB technique. In this paper, however, we concentrate our attention mainly on TLSs since they play a leading role in the processes running as a result of temperature cycling of a glass.

The potential $U(x)$ in Eqs. (1) and (2) has no linear term and this means that the origin of the coordinate $x$ is chosen at the abscissa of the extremum point of $U(x)$. In the case of single-well potentials the origin of $x$ is located at the minimum of this well, at the point ($x = 0$, $U(0) = 0$). In the case of double-well potentials, there are three extremum points: the top of the potential barrier and the minima of two adjacent wells. Therefore, a same TLS, up to an additive constant, may be described in the framework of the SAP model in three different ways, depending on the location of the origin of $x$ at one of these extremum points: (i) at the top of the barrier; (ii) at the right minimum or (iii) at the left minimum. (As usually, it is assumed that $x$ increases from left to right on the $x$-axis.) In all these cases the parameters $h$ and $x$ have differing values. The first case is realized if one considers TLSs with negative $h$ values ($h < 0$), whereas $x$ may be arbitrary, either positive or negative. The second and third cases are realized when one deals with positive $h$ values ($h > 0$), while $x$ has either positive values $x > \sqrt{32h/9}$ (in the second case) or negative values $x < -\sqrt{32h/9}$ (in the third case). Correspondingly, the distribution function $F(h,x)$ will have different forms for these differing choices.

To illustrate this multiplicity of the SAP model, we consider as an example a *symmetric* double-well potential with *equal* energies of the minima which are located at the distances $-d/2$ and $+d/2$ from the top of the barrier along the $x$ axis ($d$ denotes the total separation between the two minima of a double-well potential in general, be it symmetric or asymmetric). In our example the potential $U(x)$ of the same shape but with three different origins of the coordinates ($x,U$) is given by the following expressions:

$$U_1(x) = U_0 x^2 (x^2 - d^2/2),$$
$$U_{2,3}(x) = U_0 x^2 (x \pm d)^2. \qquad (3)$$

The physically meaningful characteristics — the height of the barrier ($V = U_0 d^4/16$) and the distance between the top of the barrier and the left or right minimum ($d/2 = (V/U_0)^{1/4}$) — are identical for these three mathematically differing forms of the same potential. However, the values of $h$ and $x$ are different: in the first case $h = -d^2/2$ and $x = 0$, whereas in the second and third cases $h = d^2$ and $x = \pm 2\sqrt{h}$.

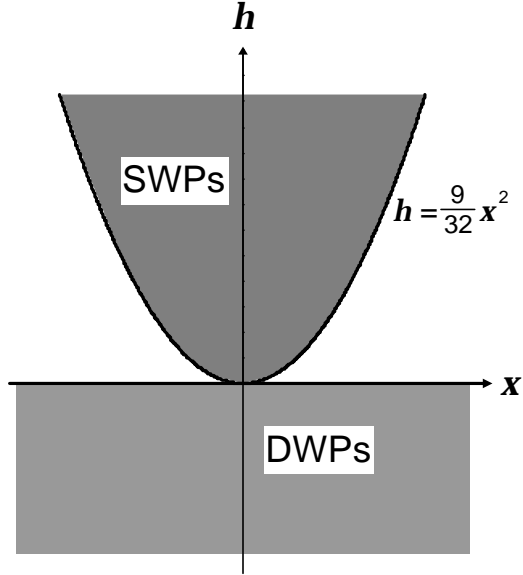

FIG. 2. Regions of different shapes of SAPs, depending on the values of parameters $h$ and $x$, shown in the ($x,h$) plane. Inside the parabola $h = 9x^2/32$ there is the domain of existence of single-well potentials (SWPs) belonging to SLMs. Half-plane $h < 0$ is the domain of double-well potentials (DWPs) describing all TLSs.

We choose the origin ($x = 0$) at the top of the barrier. In this case, *all possible* TLSs are described by Eqs. (1) and (2) with the only condition $h < 0$ to be fulfilled for both positive and negative $x$ values, while all possible SLMs are described by the condition $h > 9x^2/32$ (see Fig. 2). Our choice is best suited for description of weakly asymmetric TLSs (also called low-energy TLSs, i.e. the TLSs with a small energy difference between the minima of two wells). In Fig. 2 these TLSs are located near the vertical line $x = 0$ in the half-plane $h < 0$. For such almost symmetric TLSs with $x \cong 0$ the mean height of the barrier $V = (V_L + V_R)/2$, determined as the average of barrier heights for the left and the right well, and also the asymmetry $\Delta$ — the energy difference between the right ($x = x_R$) and the left ($x = x_L$) minimum (see Fig. 3) — can be expressed via the respective SAP parameters as follows:

$$V \approx U_0 \mathbf{h}^2/4 ; \tag{4}$$

$$\Delta \approx \sqrt{1/2} U_0 \mathbf{x} |\mathbf{h}|^{3/2} . \tag{5}$$

For later use we define the asymmetry energy $\Delta$ in Fig. 3 to be positive ($\Delta > 0$), if the right minimum $U_{\min, R}$ of the TLS's double-well potential $U(x)$ (being situated at a positive coordinate value) is located at a higher energy than the left minimum $U_{\min, L}$, i.e. $\Delta > 0$ if $U_{\min, R} - U_{\min, L} > 0$. In the opposite case, if $U_{\min, R} - U_{\min, L} < 0$, the asymmetry is considered to be negative ($\Delta < 0$).

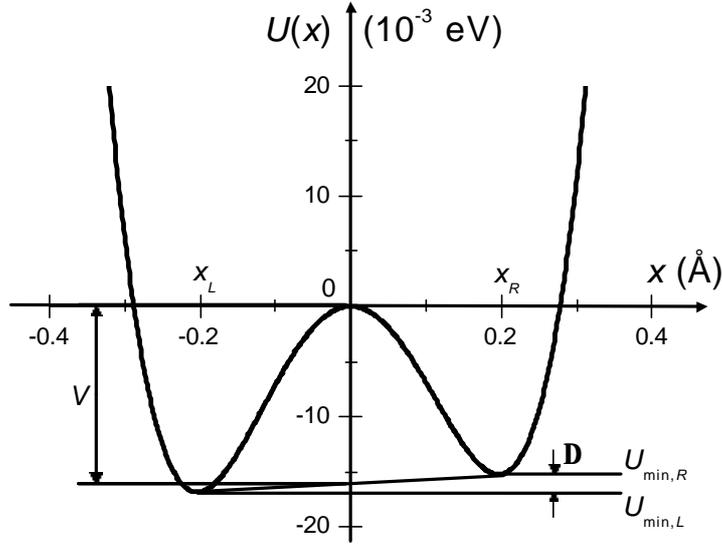

FIG. 3. A double-well potential describing a TLS with asymmetry $\Delta$ and barrier height $V$ for reasonable $\mathbf{h}$ and $\mathbf{x}$ values, $\mathbf{h} = -0.08$ and $\mathbf{x} = 0.01$, according to Eq. (1) with $U_0 = 10$ eV and $a = 1$ Å. In this case Eqs. (4) and (5) yield $V = 16$ meV and $\Delta = 1.6$ meV, so that $V/\Delta = 10$. From Eq. (1), the distance between the minima of two wells is $d \equiv x_R - x_L = 0.4001$ Å.

From Eq. (2) one can conclude that any double-well potentials (all of them having $\mathbf{h} < 0$) with positive $\mathbf{x}$ values ($\mathbf{x} > 0$) must have a positive asymmetry ($\Delta > 0$), and *vice versa*, all the potentials with any negative $\mathbf{x}$ value ($\mathbf{x} < 0$) have a negative asymmetry ($\Delta < 0$). (The same conclusion follows also from Eq. (5) because, as is seen, both $\mathbf{x}$ and $\Delta$ have the same sign.)

TLSs with small asymmetry $\Delta$ in comparison with the barrier height $V$ have a leading role in the irreversible broadening of spectral holes studied in this paper. In such a case, as will be shown later, the asymmetry parameter $\Delta$ is subjected to remarkable changes already under moderate pressures of some kilobars, giving rise to the pressure effects observed.

## B. Distribution function $F(h, x)$

An exact analytical expression for the distribution function $F(h,x)$ of random parameters $h$ and $x$ in glasses is not known. Proceeding from general considerations, it was supposed that this distribution for the harmonic force constant $h$ at a fixed value of the cubic anharmonicity $x$ has its maximum near the value $h_m \approx 1$ corresponding to the ordinary (rigid) single-well atomic potentials which are vastly prevailing in any material (either amorphous or crystalline). In contrast to these potentials, all soft anharmonic potentials (SAPs) in a glass (both TLSs and SLMs) are represented by various values of $h$ belonging to the tail of the distribution function $F(h, x = \text{const})$ at $|h| \ll 1$, with positive and negative $h$ values not far from $h = 0$ (see, e.g., Fig. 3(a) of Ref. 27).

### 1. The sign of the configurational coordinate x

In the original version of the SAP model[4], it was taken that the sign of the coordinate $x$ (the direction of its increase or decrease) may be selected arbitrarily because "there are no preferred directions in a glass and the potentials $U(x)$ with asymmetries of opposite signs are equally probable". For that reason the distribution function $F(h = \text{const}, x)$ for fixed $h$ and varying $x$ was assumed to be an even function of $x$, i.e. $F(h,x) = F(h,-x)$. In case when the origin ($x = 0$, $U(x) = 0$) is placed at the top of the barrier, the statement $F(h,x) = F(h,-x)$ means that the number of TLSs with positive asymmetry ($\Delta > 0$) should be equal to the number of TLSs with negative asymmetry ($\Delta < 0$). Probably all the authors up to the present have agreed with this assumption.

However, in our paper[28] (see also Refs. 11-13) it was noted that, in fact, there exists a possibility for a justified selection of the sign of the coordinate $x$ — one can relate the sign of the coordinate variation to the sign of the change of the glass volume, so that a decrease / increase of the coordinate $x$ would correspond to a decrease / increase of the volume. In this definition, it is taken into account that the glass volume $V$ depends on $x$. The existence of this dependence follows, e.g., from the occurrence of linear in pressure effects in glasses at low temperatures in the region of small $P$ values. It is possible to describe these effects in the framework of the SAP model if one considers the parameters of the SAP to be linearly dependent on $P$ (see, e.g., Eq. (2) of Ref. 29). Namely, for a single TLS in the low pressure limit the pressure dependence of the energy difference between the minima of the two wells $\Delta$ may be written as $\Delta(P) \cong \Delta + P\delta V$, where $\delta V$ is the difference in the glass volume for these two minima. From such a linear dependence of $\Delta$ on $P$ it is evident that the glass volume $V$ is different for the two minima of the same TLS. But since these minima are located at two different values of the coordinate $x$, one can conclude that there exists a dependence of $V$ on $x$.

### 2. Dissymmetry of compression and expansion

As we just noted, in the original SAP model[4] the distribution function of the cubic anharmonicity parameter *x* was taken to be symmetric, $F(\mathbf{h},\mathbf{x})$ being an even function of *x*. However if *x* is defined in the above manner then one cannot any more consider that "the potentials $U(x)$ with asymmetries of opposite signs are equally probable". This should mean that the isotropic compression in glasses is equivalent to isotropic stretching (expansion). In fact, however, this is not the case. As it is well known, in the strong compression limit any condensed matter becomes metallic, while in the strong expansion limit one gets a gas or plasma. Such a fundamental dissymmetry stems from asymmetry of the intermolecular potentials (e.g., Lennard–Jones potential) already. The compression – expansion symmetry can exist only in the harmonic approximation; beyond this approximation (which is certainly the case with the SAP model), such kind of symmetry does not exist. This inequivalence of compression and expansion is a fundamental property of any condensed phase, including glass state. Our definition of the sign of *x* enables to take into account this property of glass and we see that the compression – expansion dissymmetry leads to the asymmetry of the distribution function under consideration $F(\mathbf{h} = \text{const}, \mathbf{x})$ with respect to the sign of *x*. This dissymmetry must be taken into account when applying the SAP model with the aim to describe the effects related to changes of the glass volume.

In our study of SLMs under pressure[11] it was demonstrated that for the given definition of the sign of *x* the SLMs with a small positive **h** value and small *negative* **x** values turn out to be more abundant than SLMs with the same **h** value but small *positive* **x** values, so that the maximum of $F(\mathbf{h},\mathbf{x})$ for a small positive **h** value is located at $\mathbf{x}_m < 0$ and not at $\mathbf{x}_m = 0$. (It is notable that the usual pair potentials for atoms and molecules also have a negative cubic anharmonicity parameter **x**.)

### 3. Distribution function at small |**h**| and |**x**|

In this paper we will show that actually the sign of the asymmetry of the distribution function $F(\mathbf{h},\mathbf{x})$ with respect to the sign of **x**, i.e. whether $F(\mathbf{h},\mathbf{x})$ has a maximum at some negative or positive **x** value, depends on the sign of **h**. Namely, if **h** > 0, as it is for all SLMs, then $\mathbf{x}_m < 0$. However, if **h** < 0, then it turns out that $\mathbf{x}_m > 0$. To show this we take into account that a SAP $U(x)$ given by Eq. (2) is in fact a sum of a number (*z*) of pair potentials $\Phi_j(x - x_j)$ describing interactions between the atom(s) under consideration having the coordinate *x* (the "central atom(s)") and surrounding atoms with positions $x_j$. Thus we have $U(x) = \sum_{j=1}^{z} \Phi_j(x - x_j)$. To remain within the scope of the SAP model, we need to take into account in this potential only the terms up to the 4th power of *x* inclusive.

Let us consider at first a SAP with **x** = 0. Then from Eq. (2) we have $U(x) = U_0(\mathbf{h} x^2 + x^4)$. We can choose all $x_j$ in such a way that the value **x** = 0 would be realized at the points $\{x_j = 0\}$. Now the condition of absence of the linear in *x* term in $U(x)$ reads $\sum_j \Phi'_j(0) = 0$, whereas the harmonic force constant $h$, which is equal to the second derivative of the potential $U(x)$ with respect to the coordinate, is expressed as $h = \sum_j \Phi''_j(0)/2U_0$. Further we consider another SAP, denoted as $U^*(x)$, with the atoms $j$ being slightly shifted from their positions in the first SAP. The coordinate of the minimum of $U^*(x)$ and the value of $x$ in the latter SAP are also slightly different from the respective coordinate and the value $x = 0$ for the initial SAP. One can find these differences in the following way. At small $x_j$ we can expand $U^*(x)$ into a series in terms of $x_j$ and restrict ourselves to linear terms only. Then we get $U^*(x) = U(x) + bx$, where $b = -\sum_j \Phi'_j(0) x_j$. This expression, up to a small additive constant, can be presented in the form $U^*(x) = U_0 (h\, x^{*2} + x^* x^{*3} + x^{*4})$, where $x^* = x - x_0$. Here $x_0 = \sum_j \Phi'_j(0) x_j / 2U_0 h$ is the shift of the origin, while $x^* = 4x_0 = 2\sum_j \Phi'_j(0) x_j / U_0 h$ is the new value of $x$. From this consideration it follows that the probability of finding various SAPs with a small variable cubic anharmonicity parameter $|x|$ at a fixed value of the harmonic force constant $h$ can be written as

$$F(h,x) \cong \int \cdots \int d\left(x - 2\sum_j \frac{\Phi''_j(0) x_j}{U_0 h}\right) \overline{P}(\{x_j\}) \prod_j dx_j \ , \quad (6)$$

where $\overline{P}(\{x_j\})$ is the probability of finding a set of atoms with the coordinates $\{x_j\}$ (this probability has no peculiarities).

Depending on the magnitude of $h$, the cases of small and large $|h|$ values should be distinguished here.

In the small $|h|$ limit [when $|h| \ll z\overline{\Phi''_j}(0)/2U_0$, where $\overline{\Phi''_j}(0)$ is the mean value of $\Phi''_j(0)$] the actual $h$ value is the sum of a number of random terms $\Phi''_j(0) x_j / 2U_0$ having either positive or negative sign and partially compensating each other. In this case, one can satisfy the relation $h = \sum_j \Phi''_j(0)/2U_0$ for very different $\{x_j\}$ and $\Phi''_j(0)$ values, including those for which $|\Phi''_j(0)| \gg 2U_0|h|/z$. Therefore the quantities $h$ and $\Phi''_j(0)$ in Eq. (6) become practically independent. As a consequence, for small $|h|$ and $|x|$ values one can show (bearing in mind the following property of the δ-function: $\delta(ax) = \delta(x)/|a|$, where $a$ = const) that the distribution function $F(h,x)$ has the $|h|$-type (or "seagull") singularity[5,6] and its dependence on $x$ occurs only via the product of $h$ and $x$[28], i.e. in this case the values of the parameters $h$ and $x$ are correlated. For small $|hx|$ values the $x$ dependence of $F(h,x)$ may be approximated by a linear function. As a result one gets[11-13]

$$F(h, x) \propto |h| (1 - bhx). \quad (7)$$

Here ***b*** is some parameter, having a value of the order of 1, and the linear in ***x*** term, ***bhx***, depicts the asymmetry of the distribution function with respect to ***x***. In our previous papers[11,12] it was shown that the pressure effect on SLMs can be correctly described if one supposes that the parameter ***b*** is positive (***b*** > 0). This sign of ***b*** corresponds to a *negative* asymmetry in ***x*** distribution ($x_m < 0$) for single-well potentials (as we know, these potentials describe SLMs and they have positive ***h***). However, for double-well potentials (which describe TLSs and always have negative ***h***) the positive sign of ***b*** corresponds to a *positive* asymmetry in ***x*** distribution ($x_m > 0$). This allows us to assert the following: *in glasses the low-energy TLSs with the right well (located at $x > 0$) having a higher energy than the left one (i.e. TLSs with $\Delta > 0$) should be more numerous than the TLSs with $\Delta < 0$ where the left well (with $x < 0$) has a higher energy than the right one*. Below we shall show that this statement is in full agreement with our experimental data on the effect of pressure on TLSs.

Note that the obtained dependence of the distribution function $F(\mathbf{h},\mathbf{x})$ on ***h*** and ***x*** given by Eq. (7) is in good conformity with experiment, at least for small positive ***h*** values belonging to SLMs. Indeed, as was shown in Ref. 5, existence of the factor |***h***| allows one to explain the $w^4$ power dependence of the DOS of SLMs at small ***w*** (this $w^4$ dependence is responsible for an increase in the heat capacity $C_P(T)/T^3$ from the minimum value at $T_{\min} = 0.5 \div 4.0$ K up to its peak at $T_{\max} = 2.7 \div 13.5$ K observed for various organic and inorganic glasses;[30-32] in particular, for glassy polystyrene[33] $T_{\min} = 0.9$ K and $T_{\max} = 3.0$ K). It also allows us to explain the $\propto T^{3/2}$ dependence of the irreversible increase in the spectral hole width on the cycling temperature, observed in temperature cycling experiments at ambient pressure as well as at elevated pressures (see Sec. III C and III D, respectively).

The term in the form $\propto |\mathbf{h}|\mathbf{hx}$ appearing in Eq. (7) and describing the asymmetry of the distribution function $F(\mathbf{h}, \mathbf{x})$ with respect to ***x***, has an important consequence: it ensures that the $w^4$ dependence of the DOS of SLMs at small ***w*** values remains valid under moderately high pressures as well.[11,28] (If one would take this term, e.g., in the form $\propto |\mathbf{h}|\mathbf{x}$, then the $w^4$ power dependence of the DOS of SLMs at small ***w*** values under pressure would be replaced by the $Pw^2$ dependence and this would lead to qualitative changes of the low-temperature properties of glass at rather low pressures already.)

### *4. Distribution function of V and* $\mathbf{\Delta}$

Equation (7) allows us to find the distribution function for another set of parameters which may be equally used to describe SAPs, namely the distribution $\overline{F}(V,\Delta)$ for the barrier height *V* in the case $|\Delta| << V << U_0$, and for the asymmetry energy $\Delta$. The distributions $F(\mathbf{h},\mathbf{x})$ and $\overline{F}(V,\Delta)$ are related to each other in the following manner:

$$\overline{F}(V,\Delta) = F(\mathbf{h},\mathbf{x}) \left| \frac{\partial(\mathbf{h},\mathbf{x})}{\partial(V,\Delta)} \right|, \tag{8}$$

where $h \approx -2(V/U_0)^{1/2}$ and $x \approx (\Delta/2) V^{-3/4} U_0^{-1/4}$, as follows from Eqs. (4) and (5), whereas $|\partial(h,x)/\partial(V,\Delta)| \equiv |(\partial h/\partial V)(\partial x/\partial \Delta) - (\partial h/\partial \Delta)(\partial x/\partial V)|$ is the respective transformation Jacobian. Substituting here the relation for $F(h,x)$ according to Eq. (7), $F(h,x) \propto |h|(1 - bhx)$, and using the above expressions for $h$ and $x$ as well as the relation for the Jacobian $|\partial(h,x)/\partial(V,\Delta)| \approx (2 V^{5/4} U_0^{3/4})^{-1}$, which follows from the same expressions for $h$ and $x$, one gets

$$\overline{F}(V,\Delta) \propto \frac{1 + b\Delta/(VU_0^3)^{1/4}}{(V^3 U_0^5)^{1/4}}. \tag{9}$$

Note that this equation is valid only for small $\Delta$ values, when the term $b\Delta$ is small as compared to $(VU_0^3)^{1/4}$, i.e. when $b\Delta/(VU_0^3)^{1/4} \ll 1$. One sees from Eq. (9) that in the case of positive $b$ the distribution of TLSs with respect to the sign of $\Delta$ is *asymmetric* — there are more TLSs with positive $\Delta$ values than those with negative $\Delta$ values. As already explained above, this asymmetry stems from dissymmetry between the decrease and increase in the glass volume.

## 5. Domain of applicability of Eq. (7)

Above we considered the case of small $|h|$ values. Now, in the case of large $|h|$, the most probable values of $\Phi_j''$ are close to the mean value $\overline{\Phi_j''}(0) \approx 2U_0 h/z$, because in this case $h = \sum_j \Phi_j''(0)/2U_0 \approx z\overline{\Phi_j''}(0)/2U_0$. Substituting this value $\overline{\Phi_j''}(0) \approx 2U_0 h/z$ into Eq. (6) one obtains that the factor $1/U_0 h$ under the δ-function cancels out and the δ-function no longer depends on $h$. This means that the values of $h$ and $x$ in the distribution function $F(h,x)$ become statistically uncorrelated, i.e. they are independent of each other. Consequently, in the case of large $h$ the above Eq. (7) does not work any more, because, as we found, it was applicable only for correlated $h$ and $x$, where both $h$ and $x$ were rather small ($|h| \ll 1$; $|x| \ll 1$) and the product $|hx|$ had a small value as well.

The domain of applicability of Eq. (7) is limited by the following $x$ and $h$ values: $|x| \ll 1$ and $-h_{\lim} < h < h_{\lim}$. The value of $-h_{\lim}$ can be estimated from the position of the so-called relaxation peak[6] which corresponds to the cutoff energy on the high-energy side of the distribution function $F(h = -2(V/U_0)^{1/2}, x = 0)$ for barrier heights $V$. In vitreous silica, for instance, this cutoff for the barrier energies, $V_c$, was estimated to be about $k_B \times 750$ K (see point 3.2.2 of Ref. 6). From Eq. (4) we have $h_{\lim} \approx \pm 2\sqrt{V_c/U_0}$, which yields $|h_{\lim}| \approx 0.15$, as marked in Fig. 4. We emphasize that, according to our consideration, the value of $h_{\lim}$ is determined by fluctuations in the local structure of glass. Note also another opinion[5,6] that the value of $h_{\lim}$ from the positive side of $h$ is determined by the interaction strength between different SLMs.

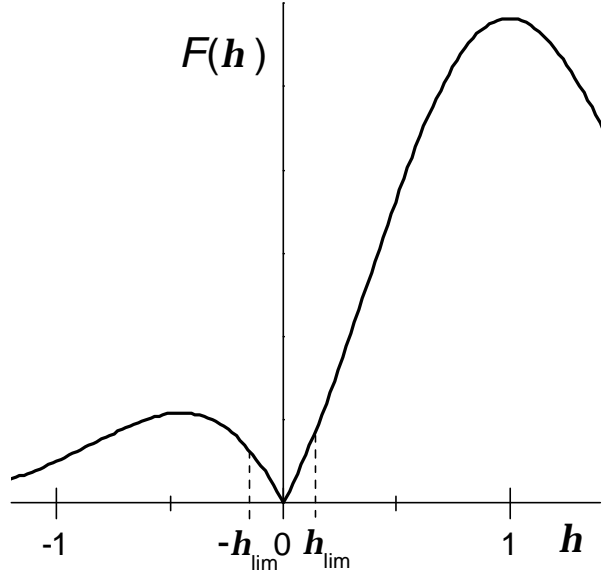

FIG. 4. Sketch of the distribution function $F(h,x)$ for $h$ at small fixed $x$ values ($|x| \ll 1$). The boundaries of variation for positive and negative values of $h$ in Eq. (7), $h_{\text{lim}}$ and $-h_{\text{lim}}$, are indicated as being estimated for vitreous silica (see the text).

## C. Residual broadening of spectral holes due to temperature cycling

As is known, an additional irreversible broadening of spectral holes in temperature cycling experiments (see, e.g., Refs. 19,34) is caused by spectral diffusion[35] during the temperature cycle, owing to presence of TLSs in a glass. We shall describe the irreversible spectral diffusion following the approach presented in the literature,[19] realizing it, however, within the framework of the SAP model and taking into account only the thermally activated processes (tunneling processes begin to play an important role at much lower temperatures, roughly at $T < 1$ K).

Let us consider first the case of normal pressure. If we assume that the asymmetry energy $\Delta$ is much smaller than the barrier height $V$ of TLSs for both positive ($\Delta > 0$) and negative ($\Delta < 0$) asymmetries, i.e. $|\Delta| \ll V$, then we can neglect the $\Delta$ dependence of the distribution function, so that we can take: $\overline{F}(V,\Delta) \approx \overline{F}(V,\Delta=0) \equiv \overline{F}(V,0)$. In this case the irreversible increase in the hole width due to thermally induced spectral diffusion only, $\Gamma_{\text{SD}}$, can be expressed as

$$\Gamma_{SD} \approx \frac{1}{2} \int_{T_b}^{T_c} dT \int d\Delta \int dV C(V,\Delta) \overline{F}(V,0) \operatorname{sech}^2\left(\frac{\Delta}{2k_B T}\right) d\left(T - T_f(V,\Delta,t)\right), \qquad (10)$$

where $T_b$ and $T_c$ are the hole burning and cycling (excursion) temperatures, respectively. Here we have neglected the contribution of the tunnel splitting $\Delta_0$ to the energy separation $E$ between the lowest energy levels in the right and left potential wells. As is known (see, e.g., Ref. 36), this energy difference $E$ (also called simply the energy of a TLS) is given by the relation $E = \sqrt{\Delta^2 + \Delta_0^2}$. The tunnel splitting $\Delta_0$ due to the coupling between these two levels can be written as $\Delta_0 = \hbar w_0 \exp(-l)$, where $\hbar w_0 = E_0$ is the ground state energy in each well (assumed to be the same for both wells), and the tunneling parameter $l = (d/\hbar)\sqrt{2mV}$ depends on the mass $m$ of the tunneling particle, the barrier height $V$ and the distance $d$ between the two minima, but not on the temperature $T$. It is evident that in the case of high potential barriers between the wells, as the temperature is increased, the tunneling transitions are playing progressively smaller role as compared to the thermally activated overbarrier jumps, since the number of these jumps grows rapidly with $T$ (namely $\propto \exp(-V/k_B T)$ according to the Arrhenius law). Indeed, it is known (see, e.g., Fig. 4.1 of Ref. 37) that for most TLSs with $E/k_B T \geq 1$, i.e. at temperatures above ~1 K, the condition $\Delta_0 << |\Delta| \cong E$ is satisfied. Therefore in temperature cycling experiments at temperatures exceeding ~1 K one can neglect the term $\Delta_0$ and use the approximation $E \cong |\Delta|$.

The obtained Eq. (10) is similar to Eq. (8) of Ref. 19, except for some differences in notation and introduction of the approximation $E \cong |\Delta|$. The physical meaning of Eq. (10) is as follows.[19] The irreversible broadening of a spectral hole arises from interaction of an electronic transition in a probe molecule with TLSs, which had irreversibly changed their state (jumped over barrier from one potential well to another) during the temperature cycle. It is assumed in Eq. (10) that this interaction is effected by way of coupling between elastic dipoles ($\propto 1/r^3$). In this case the diffusional broadening kernel has a Lorentzian shape[35,38] and its width $\Gamma_{SD}$ (FWHM) is proportional to the number of transformed TLSs. The coefficient of proportionality $C(V,\Delta)$ for small $\Delta$ values ($\Delta << V$), which describes the interaction strength between TLSs and a probe transition (takes into account the change of the frequency of an electronic transition in the probe molecule as a result of jumping over the barrier), depends linearly on the distance $d$ between the minima of a double-well potential of a TLS, i.e. $C(V,\Delta) \propto d$. On the other hand, for symmetric potentials (with $x = 0$ and hence $\Delta = 0$) we had $d = 2(V/U_0)^{1/4}$ (see Sec. III A). Therefore, for weakly asymmetric potentials ($x << 1$ and $\Delta << V$) we may write $d \approx 2(V/U_0)^{1/4}$. Consequently, we get

$$C(V,\Delta) \propto 2(V/U_0)^{1/4}. \qquad (11)$$

In deriving the above Eq. (10) for $\Gamma_{SD}$ it was also assumed that below a certain temperature $T_f = T_f(E,V,t_e)$ (the so-called freezing temperature[19]) the corresponding TLSs fall out of equilibrium, i.e. their characteristic relaxation times $t$ become longer than the experimental time $t_e$. At temperatures above $T_f$, on the contrary, a full relaxation occurs and the thermal equilibrium is reached, determined by the energy difference $E$ between the eigenstates in the two minima of a double-well potential. The factor $\frac{1}{2}\text{sech}^2(\Delta/2k_BT)$ defines the probability of changing the "well of location" for a TLS at temperature $T$. We suppose that in the actual for us temperature region 5 K < $T$ < 18 K the relaxation of TLSs occurs mainly via thermally activated (overbarrier) processes and therefore, as was mentioned above, we can take into consideration only such processes. Their rate follows the Arrhenius law $t^{-1} = t_0^{-1}\exp(-V/k_BT)$, where $t_0^{-1}$, the "frequency of attempts" to overcome the barrier of height $V$, is on the order of typical vibration frequencies of matrix, say 1 THz (or about 33 cm$^{-1}$). The SHB experiments are characterized by very long duration ($t_e$ ~ 100 s) as compared to the timescale $t_0$ ~ $10^{-12}$ s. Comparison between the time intervals $t_e$ and $t = 10^{-12}\exp(V/k_BT)$ shows that during the time $t_e$ the barriers with heights up to $V \approx 32\ k_BT$ can be surmounted, and thus under these conditions all the TLSs with barriers $V \leq 32\ k_BT$ will contribute to $\Gamma_{SD}$. At the temperatures of our experiments (between 5 and 18 K) these TLSs have barriers of heights $V \leq 32\ k_B \times (5 \div 18)$ K, or $V \leq 0.014 \div 0.050$ eV. Thus, one deals predominantly with rather "hard", high-barrier TLSs. For these TLSs we have $V/U_0 \leq 0.0014 \div 0.005$ (recall that according to Eq. (1), $U_0$ ~ 10 eV), and using Eq. (4) we can estimate the respective $h$ values: $|h| \leq 0.07 \div 0.14$.

In fact, the actual upper limit for values of $E = \sqrt{\Delta^2 + \Delta_0^2} \cong |\Delta|$ is much lower due to the factor $\frac{1}{2}\text{sech}^2(\Delta/2k_BT)$ in Eq. (10), which includes a term $\exp(-|\Delta|/2k_BT)$ with asymptotic exponential behavior and thus cuts off the more asymmetric potentials (TLSs with greater values of asymmetry $|\Delta|$). However, this upper limit is still much higher than the tunneling contribution $\Delta_0$ to $E \cong |\Delta|$, i.e. we have $\Delta_0 \ll |\Delta|$. Thus, the working TLSs in our case are moderately hard ones, and their energy is well determined by the asymmetry $\Delta$ of double-well potentials.

The condition for estimating the freezing temperature $T_f$ can be obtained by equating the characteristic time of a thermally activated process to the experimental time $t_e$, i.e. $t(E,V,T_f) \cong t(|\Delta|,V,T_f) = t_e$. For an activated process in the case of $T = T_f$, one has $t^{-1} = t_0^{-1}\exp(-V/k_BT_f)$ and the condition $t = t_e$ yields

$$T_f \cong V\,[k_B \ln(t_e/t_0)]^{-1}. \qquad (12)$$

Because the range of values for $E$ and hence also for $|\Delta|$ is much narrower than that for $V$, we have assumed here that the barrier-crossing rate is the same for particles in either of the two wells and is determined by the average of barrier heights for the left and right well: $V = (V_L + V_R)/2$.

In this approximation, by using Eqs. (9), (11), (12) and the relation $\bar{F}(\Delta,V) \approx \bar{F}(0,V)$, the integration in Eq. (10) is straightforward and yields

$$\Gamma_{SD} \propto [\ln(t_e/t_0)]^{1/2} (T_c^{3/2} - T_b^{3/2}). \tag{13}$$

The obtained dependence of the irreversible increase in hole width on the cycling temperature $T_c$ and hole burning temperature $T_b$ coincides with the corresponding functional dependence on temperature ($T^{3/2}$) described by Eq. (13) of Ref. 19 and is in good agreement with experiment.

Note that the asymmetry term $\propto \boldsymbol{b h x}$ in Eq. (7) or the equivalent term $\propto \boldsymbol{b}\Delta$ in Eq. (9) does not influence the magnitude of the integral in Eq. (10), because the contribution from this term cancels out on integration. Of course, Eq. (9) is only some approximation to the real distribution function and applies in the region of not too large values of $E \cong |\Delta|$. This is, however, sufficient for our purposes due to the "fast" cutoff term $\text{sech}^2(\Delta/2k_BT)$ in the integral of Eq. (10).

It is also essential to note that a $T^{3/2}$ dependence derived in Eq. (13) lends an experimental support to the occurrence of a "seagull" type distribution function $F(\boldsymbol{h}, 0) \propto |\boldsymbol{h}|$ for small $|\boldsymbol{h}|$ on the side of negative $\boldsymbol{h}$ values ($\boldsymbol{h} < 0$). We remind that existence of the maximum in the $C_P/T^3$ dependence for glasses provides support to this distribution function on the side of positive $\boldsymbol{h}$ values[5] ($\boldsymbol{h} > 0$).

## D. Effect of pressure on residual broadening $\Gamma_{SD}$ of spectral holes

In order to incorporate the pressure dependence into $\Gamma_{SD}$, we proceed from the idea that it can be done by assuming a linear coupling of TLSs to the pressure-induced elastic compression at small $|x|$ values.[26,28] Based on this assumption (usually used also for describing an interaction between SAPs and sound waves; see, e.g., Ref. 26), the effect of pressure $P$ on a SAP in the low pressure limit may be taken into account by adding to the potential $U(x)$ of Eq. (2) a term $U_0\boldsymbol{k}Px$, linear in $P$ and $x$. So the pressure-dependent potential, up to an inessential here additive term can be expressed as

$$U_P(x) = U_0 (\boldsymbol{h} x^2 + \boldsymbol{x} x^3 + x^4) + U_0\boldsymbol{k}Px. \tag{14}$$

The last term in this equation takes into account the $PV$-term in the free energy of glass at pressure $P$ and also the discussed above dependence of $V$ on $x$ (see Subsection III B 1). The coefficient $\boldsymbol{k}$ in Eq. (14) obeys some distribution as do the other model parameters. In consequence of the chosen definition for the sign of $x$, the quantity $\boldsymbol{k}$ is positive. Note that some SAPs may have rather small (but in any case positive) $\boldsymbol{k}$ values. However, in order to couple to external hydrostatic pressure $P$ in linear approximation a double-well potential *must* have $\boldsymbol{k} \neq 0$. The circumstance that a typical SAP has the value of $\boldsymbol{k}$ different from zero follows also from the symmetry considerations if one keeps in mind that there exist no universal symmetry properties for soft potentials in glasses. Quite the reverse, one might expect, in general, that local atomic configurations in a glass have low symmetry and hence there should be a change of the glass volume accompanying transitions in soft systems. Here we restrict our consideration to the linear in pressure effects only. These effects are described by the mean value of linear in $kP$ terms. As the result of averaging over these terms, the "running" value of $k$ is replaced by its mean value $k_{mean}$ that is a positive quantity. This allows us, for the further consideration, to use the model with a single value of this parameter, namely $k_{mean}$.

Similarly to our earlier study,[11] we assume that $k_{mean}$ is on the same order of magnitude as typical experimental values for the bulk compressibility $k_{bulk} \sim 10^{-5}$ bar$^{-1}$ of organic solids. (For simplicity we shall omit the subscript "mean" everywhere below, bearing in mind, however, that $k \equiv k_{mean}$.)

Introducing a new, shifted coordinate $X = x + kP/2h$, one can present Eq. (14), up to an additive constant (independent of $x$), in the conventional for the SAP model form:

$$U_P(X) = U_0 (h_P X^2 + x_P X^3 + X^4) + const , \qquad (15)$$

where $h_P = h + \eth h_P$ and $x_P = x + \eth x_P$ are the values of $h$ and $x$ under pressure, while $\eth h_P \approx -3xkP/2h$ and $\eth x_P \approx -2kP/h$ (including only terms linear in $P$).[11]

One can see that the main effect of pressure in the actual region of moderately hard (high-barrier) TLSs with almost equally deep wells ($x \approx 0$) lies in the change of the asymmetry energy $\Delta$ for these wells: $\Delta_P = \Delta + \eth\Delta$, where $\eth\Delta = U_0 kP d \cong (2|h|)^{1/2} U_0 kP$ ($d$ is the distance between the minima of two wells, see Fig. 3). Since the last term in Eq. (14) is positive/negative for any positive/negative value of the coordinate $x$ at a fixed pressure $P$ (because $k$ and $P$ are both positive quantities), the energy $U_{min, R}$ of the right minimum with $x = x_R > 0$ in the double-well potential of a TLS raises with increasing pressure, whereas the energy $U_{min, L}$ of the left minimum with $x = x_L < 0$ at the same time falls with respect to the top of the barrier at $x = 0$ (for notations, see Fig. 3). This means that initially symmetric or nearly symmetric TLSs (which had $\Delta \cong 0$ at normal pressure) transform under pressure $P$ to asymmetric TLSs with a positive asymmetry ($\Delta_P > 0$) that is growing with increasing pressure. For spectral holes at low temperatures, these TLSs fall out of play in the sense that they do not contribute to the residual broadening $\Gamma_{SD}$ of holes, because the contribution to $\Gamma_{SD}$ comes only from those TLSs, which are at pressure $P$ almost symmetric ($\Delta_P \cong 0$).

On the other hand, a certain number of TLSs which had before the pressure application a negative asymmetry ($\Delta < 0$), transform under the same pressure $P$ to symmetric or nearly symmetric TLSs with $\Delta_P \cong 0$. Thus, they come into play instead of the TLSs, which had $\Delta \cong 0$ at normal pressure. However, Eq. (9) shows that TLSs with a small negative asymmetry ($\Delta < 0$) are always less abundant than those with a near-zero or small positive asymmetry ($\Delta \geq 0$), since according to Eq. (9) we have for the distribution function $\overline{F}(V,\Delta)$ the following inequality: $\overline{F}(V,\Delta < 0) < \overline{F}(V,\Delta \geq 0)$. Therefore, as a net result of applying the external pressure, the number of nearly symmetric TLSs with $\Delta_P \cong 0$ must decrease with increasing pressure. But this in its turn means that $\Gamma_{SD}$ must decrease under pressure, because relevant are only those TLSs which at high pressure have almost equally deep wells with $|\Delta_P| \leq k_B T$ (with $|\Delta_P| \leq 1.6$ meV for $T_c \leq 18$ K in our temperature cycling experiments).

For describing this pressure effect quantitatively, one needs to find the distribution function $\overline{F}(V,\Delta)$ at elevated pressure $P$ for small $|\Delta|$ values, i.e. the distribution $\overline{F}_P(V_P, \Delta_P \approx 0)$. Under pressure, in the case $\Delta_P \approx 0$ we have also $\mathbf{x}_P \approx 0$, whereas for $\mathbf{h}_P$ we get from Eq. (4) the value $\mathbf{h}_P \approx -2\sqrt{V_P/U_0}$ (since $\mathbf{h}_P$, similarly to $\mathbf{h}$, is always negative for TLSs). According to Eq. (8), we may write

$$\overline{F}_P(V_P, \Delta_P \approx 0) \approx F_P(\mathbf{h}_P = -2\sqrt{V_P/U_0}, \mathbf{x}_P = 0) \left| \frac{\partial(\mathbf{h}_P, \mathbf{x}_P)}{\partial(V_P, \Delta_P)} \right|, \qquad (16)$$

where $F_P(\mathbf{h}_P, \mathbf{x}_P)$ is the distribution function for the parameters $\mathbf{h}_P$ and $\mathbf{x}_P$ under pressure $P$, while the Jacobian is given by the expression $|\partial(\mathbf{h}_P, \mathbf{x}_P)/\partial(V_P, \Delta_P)| \approx (2\, V_P^{5/4} U_0^{3/4})^{-1}$, obtained in the same manner as the respective expression at normal pressure for the distribution function $\overline{F}(V, \Delta)$ in Eq. (8).

To find the distribution function $F_P(\mathbf{h}_P, \mathbf{x}_P)$ in Eq. (16), we make use of the following relation:

$$F_P(\mathbf{h}_P, \mathbf{x}_P) = F(\mathbf{h}, \mathbf{x}) \left| \frac{\partial(\mathbf{h}, \mathbf{x})}{\partial(\mathbf{h}_P, \mathbf{x}_P)} \right|. \qquad (17)$$

Here $\mathbf{h} = \mathbf{h}_P - \delta\mathbf{h}_P$ and $\mathbf{x} \approx \mathbf{x}_P - \delta\mathbf{x}_P$ are the values of $\mathbf{h}$ and $\mathbf{x}$ they had before pressure application, whereas $\delta\mathbf{h}_P \approx -3\mathbf{x}kP/2\mathbf{h}$ and $\delta\mathbf{x}_P \approx -2kP/\mathbf{h}$ (see above). Taking into account that under pressure $\mathbf{x}_P \approx 0$, we get $\mathbf{x} \approx -\delta\mathbf{x}_P$ and $\delta\mathbf{h}_P \approx -3(kP/\mathbf{h})^2$. The latter quantity is quadratic in $P$ and therefore very small at low pressures, so that it may be taken equal to nearly zero: $\delta\mathbf{h}_P \approx 0$. As a result we get that before the pressure application (i.e. at $P = 0$) the TLSs under consideration (with $\Delta_P \approx 0$) had the parameters $\mathbf{h} \approx \mathbf{h}_P$, $\mathbf{x} \approx 2kP/\mathbf{h}$ (or $\mathbf{x} \approx -2kP/|\mathbf{h}|$, since $\mathbf{h}$ is negative for all TLSs), $V \approx V_P$, and $\Delta \approx -2kPV^{1/4} U_0^{3/4}$ (the last two relations were obtained by means of Eqs. (4) and (5)). In this approximation, where the terms higher than linear in $P$ are neglected, the transformation Jacobian in Eq. (17) is equal to 1 and therefore Eq. (17) gives us $F_P(\mathbf{h}_P, \mathbf{x}_P \approx 0) \approx F(\mathbf{h}, \mathbf{x} \approx -2kP/|\mathbf{h}|)$.

Further, according to Eq. (7) we get $F(\mathbf{h}, \mathbf{x} \approx -2kP/|\mathbf{h}|) \approx (1 - 2\mathbf{b}kP)\, F(\mathbf{h}, \mathbf{x} = 0)$, where $F(\mathbf{h}, \mathbf{x} = 0) \propto |\mathbf{h}|$ from Eq. (7). Thus, we have $F_P(\mathbf{h}_P, \mathbf{x}_P \approx 0) \approx (1 - 2\mathbf{b}kP)\, F(\mathbf{h}, \mathbf{x} = 0)$. Inserting this relation for $F_P(\mathbf{h}_P, \mathbf{x}_P \approx 0)$ into Eq. (16), we obtain $\overline{F}_P(V_P, \Delta_P \approx 0) \approx (1 - 2\mathbf{b}kP) \times |\mathbf{h}| \times (2V_P^{5/4} U_0^{3/4})^{-1}$, where $|\mathbf{h}| \approx 2(V/U_0)^{1/2}$ from Eq. (4), while $(2V_P^{5/4} U_0^{3/4})^{-1}$ is the value of the Jacobian in Eq. (16). Bearing in mind that $V \approx V_P$, we get

$$\overline{F}_P(V_P, \Delta_P \approx 0) \approx (1 - 2\mathbf{b}kP)\, \overline{F}(V, \Delta = 0), \qquad (18)$$

where $\overline{F}(V, \Delta = 0) \propto (V^3 U_0^5)^{-1/4}$ according to Eq. (9).

From Eq. (18) it is evident that external pressure applied to the sample reduces the number of the low-energy (i.e. weakly asymmetric) TLSs in a glass if $\mathbf{b}$ is a positive quantity.

Finally, one obtains the following pressure dependence of the irreversible increase in the hole width due to temperature cycling:

$$\Gamma_{SD}(P) \approx A\, (1 - 2\mathbf{b}kP)\, [\ln(t_e/t_0)]^{1/2}\, (T_c^{3/2} - T_b^{3/2}), \qquad (19)$$

where $A$ is a constant (independent of $P$ and $T$). Thus, within the limits of used approximations, application of pressure results in a uniform (over the whole temperature range between $T_b$ and $T_c$) reduction of the irreversible broadening of a spectral hole due to thermally induced spectral diffusion, and this reduction is proportional to $P$. It should be emphasized that the $\propto T^{3/2}$ dependence of $\Gamma_{SD}$ in our theory is maintained at moderately high pressures; this is so owing to the specific $\propto |h|hx$ form of the asymmetric with respect to $x$ term in the derived distribution function $F(\mathbf{h}, \mathbf{x})$ given by Eq. (7).

## IV. EXPERIMENTAL RESULTS AND DISCUSSION

In Fig. 5 are summarized the obtained results on the residual broadening $\Gamma_{SD}$ of spectral holes at various fixed pressures. Here, each experimental point represents the total hole width $\Gamma_{hole} = \Gamma_0 + \Gamma_{SD}$, recorded after completing the respective temperature cycle $T_b \rightarrow T_c \rightarrow T_b$ ($\Gamma_0$ is the initial hole width, i.e. the width of a hole burned and measured at temperature $T_b$ before the very first temperature cycle; $T_b$ is equal to 5 K or, more exactly, has slightly different values between 4.7 and 5.0 K in individual experiments). Curves are the nonlinear least-squares fits to the data points, assuming for $\Gamma_{SD}$, according to Eq. (19), the following temperature dependence: $\Gamma_{SD}(T) = \Gamma_{hole}(T) - \Gamma_0 \propto (T_c^{3/2} - T_b^{3/2})$.

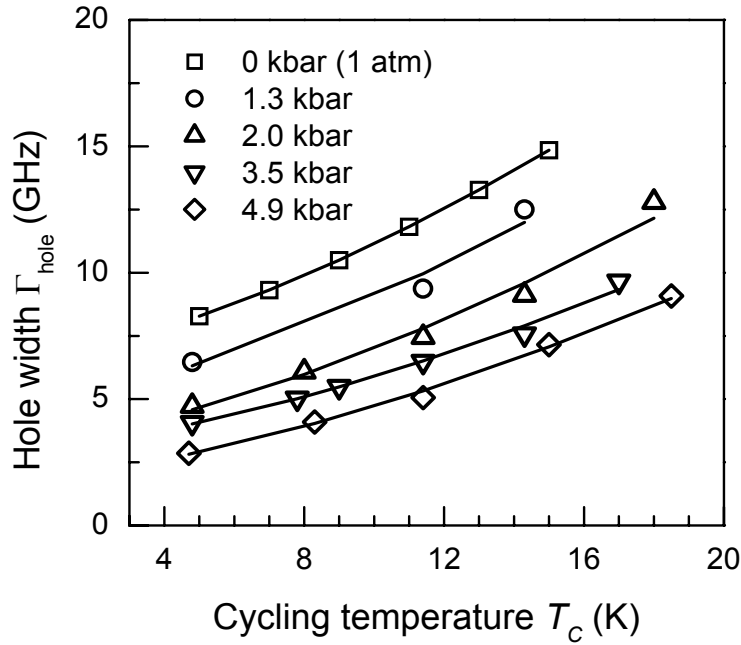

FIG. 5. Total hole width $\Gamma_{hole} = \Gamma_0 + \Gamma_{SD}$ as a function of the cycling temperature $T_c$ at different pressures ($P$ = 0, 1.3, 2.0, 3.5, and 4.9 kbar from top to bottom, respectively) for chlorin-doped glassy polystyrene. Curves are the least-squares fits to the data presented in the coordinates $T_c$ and $\Gamma_{SD}(T_c) \propto (T_c^{3/2} - T_b^{3/2})$, where $T_b = 4.7 \div 5$ K.

From Fig. 5 two main conclusions can be drawn. First, the initial width of holes burned at temperature $T_b$ under different pressures is the smaller the higher is pressure. This once more lends support to our experimentally observed and theoretically explained finding that the width of spectral holes and hence also the homogeneous width of respective zero-phonon lines in glasses decreases with increasing pressure.[9,11-13]

Second, the experimental data fit rather well to the $T^{3/2}$ dependence on the cycling temperature $T_c$ at all pressures used. Note that a similar dependence was obtained earlier for chlorin molecules in a glassy benzophenone at normal pressure.[20] Notice also that no temporal broadening of holes was observed by us when monitoring at various cycling temperatures $T_c$ over time intervals comparable to a cycle duration. This justifies neglecting the weak time dependence of the logarithmic factor in Eq. (19), which may therefore be considered as a constant.

The most important experimental result of the present study lies in the fact that the quantity $\Gamma_{SD}$, residual irreversible hole broadening due to temperature cycling, decreases with increasing pressure. It means that the number of the low-energy TLSs (i.e. TLSs with a small positive or negative asymmetry energy $\Delta$), which determine the residual broadening of holes, decreases under pressure. This result is in complete agreement with our theoretical treatment of the effect, as is obvious from Eq. (19). The obtained result is graphically illustrated by Fig. 6 where the values of the fitting coefficient

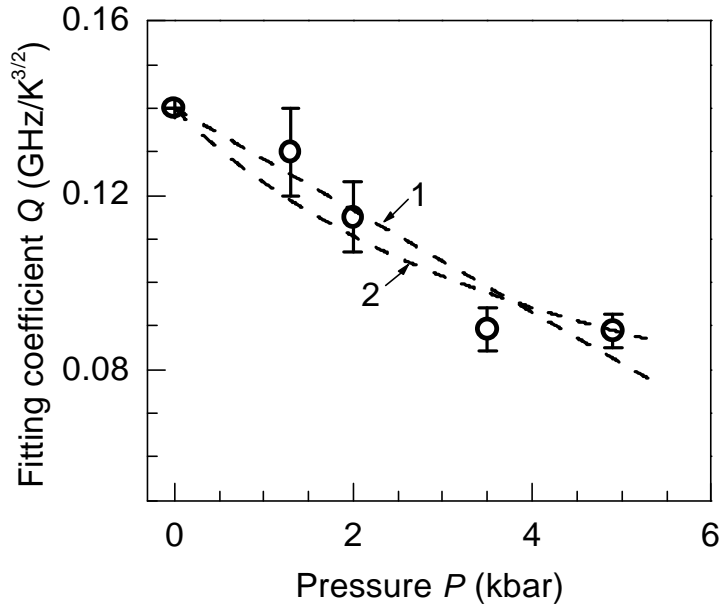

FIG. 6. Pressure variation of the values of the fitting coefficient $Q = A(1 - 2bkP)[\ln(t_e/t_0)]^{1/2}$ appearing in Eq. (19), rewritten as $\Gamma_{SD} \approx Q(T_c^{3/2} - T_b^{3/2})$, for chlorin molecules in a glassy polystyrene. Dashed lines show the least-squares fits to the data points $Q(P)$: line 1 – linear fit, if $k$ = const; line 2 – nonlinear fit, if $k = (B_0 + B_0'P)^{-1}$ (see the text).

$$Q = A \,(1 - 2\bm{b}\bm{k}P)\,[\ln(\bm{t}_e/\bm{t}_0)]^{1/2} \qquad (20)$$

for the $T^{3/2}$ dependence of $\Gamma_{SD}$ in Eq. (19) at several different pressures are shown. In fact, this fitting coefficient characterizes the slope of the individual curves in Fig. 5. One can see from Fig. 6 that, for instance, at 4.9 kbar the coefficient $Q$ is by about 36 % smaller than it was at atmospheric pressure. In other words, at a pressure of 4.9 kbar the residual hole broadening makes up only about two thirds of its initial value obtained at ambient pressure.

By using the linear regression line $Q(P) = Q_0 + (dQ/dP)P$ (see dashed line 1 in Fig. 6), where $Q_0 = A\,[\ln(\bm{t}_e/\bm{t}_0)]^{1/2} = 0.140 \pm 0.006$ GHz K$^{-3/2}$ and $dQ/dP = -2\bm{b}\bm{k}Q_0 = -0.012 \pm 0.002$ GHz K$^{-3/2}$ kbar$^{-1}$, we can estimate the magnitude of the constant $\bm{b}$, an important parameter of the theory, which appears in Eqs. (7), (9), (18), (19), (20). Taking for amorphous atactic polystyrene at $T = 4.2$ K the value of the adiabatic bulk modulus $B_S = 5.6 \times 10^3$ MPa = 56 kbar, or the bulk compressibility $\bm{k}_{\text{bulk}} \equiv 1/B_S = 0.018$ kbar$^{-1}$ (see Table 7 in the Appendix of Ref. 39), we get from the relation $Q(P) = 0.140\,(1 - 0.036\,\bm{b}\,P)$ according to Eq. (20) that the best fit value for $\bm{b}$ equals $2.3 \pm 0.3$. This seems to be quite a reasonable estimate which is consistent with the value $\bm{b} \sim 1$ assessed in Ref. 11 from the experiments of another kind (pressure dependence of the spectral hole widths).

In the model developed above, it was assumed that all the pertinent parameters of the model are independent of pressure in our pressure range. This seems to be a rather good approximation, except for the compressibility. It is known that, in general, the bulk compressibility of solids (defined as $\bm{k}_{\text{bulk}} = -(1/\bm{V})\,(\partial\bm{V}/\partial P)_T \cong -(1/\bm{V}_0)\,(\partial\bm{V}/\partial P)_T$, where $\bm{V}_0$ is the initial volume at $P = 0$ and at a given $T$) decreases with increasing pressure at a fixed temperature and also decreases with decreasing temperature at a fixed pressure, usually at 1 atm (see, e.g., Ref. 40). Unfortunately, we could not find from the literature reliable data about the pressure variation of the compressibility for polystyrene at liquid helium temperatures, although very accurate data[41] are available for various higher temperatures between 25 and 75°C.

As is seen from Fig. 6, the pressure dependence of the fitting coefficient $Q$ seems to deviate from the linear one at higher pressures. We suppose that this deviation may be caused, at least in part, by decrease of the bulk compressibility according to the relation $\bm{k}_{\text{bulk}} = 1/(B_0 + B_0'P)$, since it is known (see, e.g., Refs. 41,42) that in quite a good approximation the bulk modulus of solids increases linearly with increasing pressure: $B \cong B_0 + B_0'P$, where $B_0$ is the initial value of $B$ at ambient pressure ($P = 0$) and $B_0'$ is the derivative $(dB/dP)_{P=0}$.

Taking the expression for $Q(P)$ in the form $Q = Q_0\,[1 - 2\bm{b}P/(B_0 + B_0'P)]$, where $B_0$ and $Q_0$ have the fixed values $B_0 = 56$ kbar[39] and $Q_0 = 0.14$ GHz K$^{-3/2}$, and using for $B_0'$, in default of anything better, the fixed value $B_0' = 11.0$, determined for polystyrene in the pressure range 0 – 10 kbar at 25°C,[41] we get the nonlinear regression line shown in Fig. 6 (dashed line 2), where the best fit value for the variable $\bm{b}$ is equal to $\bm{b} = 4.1 \pm 0.4$. We think that of two estimates for the parameter $\bm{b}$ obtained above, 2.3 and 4.1, the latter one ($\bm{b} \sim 4$) is presumably more realistic.

The main ideas and model used in this paper for describing pressure effects in our temperature cycling measurements were earlier developed in order to explain the observed narrowing of spectral holes with increasing pressure in isothermal-isobaric experiments where neither temperature nor pressure was varied *between* hole burning and measuring procedures.[9-13] It was assumed that the main factor governing the width of spectral holes in these experiments was the interaction of an electronic transition in a probe molecule with soft localized vibrational modes (SLMs) which exist, according to the SAP model, only at small and positive ***h*** values (***h*** > 0). In this "soft" region of parameters ***h*** and ***x***, external pressure may exert essential influence on the *shape* of potentials $U(x)$ and it gives rise to mutual transformations of SLMs and TLSs. In order to suppress the density of states for SLMs, the *negative* asymmetry of the ***hx***-distribution is needed. This means that the product of ***h***, which is always positive for SLMs, and of ***x**_m*, the most probable value of ***x*** for SLMs, must have the negative sign, and therefore ***x**_m* must also be negative (***x**_m* < 0). Only in that case a process of transformation of SLMs into TLSs prevails over the opposite process, and the needed reduction of the DOS for low-energy excitations in a glass can be achieved. (We emphasize here once more that at sufficiently small positive and negative values of ***h*** and ***x*** around zero, say at −0.2 < ***h*** < 0.2 for ***h***, the distribution function $F(\mathbf{h},\mathbf{x})$ depends on ***x*** only via the product ***hx***.)

It should be noted that the experimental results[43,44] on relaxation of the specific volume of glasses after a fast quenching from the temperature close to $T_g$ (decrease of the specific volume in the course of relaxation, connected with changes in the DOS of TLSs), also require for their explanation the dependence of the glass volume ***V*** on the coordinate *x* of TLSs and a positive ***b*** value, i.e. a negative ***hx***-asymmetry. Indeed, as was shown in Subsection III B *3* by deriving Eq. (7), in the case of TLSs (which always have ***h*** < 0), the negative ***hx***-asymmetry needs that ***x**_m* must be positive (***x**_m* > 0), and this means that TLSs with the right minimum (corresponding to a larger volume of glass), being located at a higher energy than the left minimum, are more numerous than TLSs which, on the contrary, have the right minimum at a lower energy than the left one. In this case TLSs make a contribution to the thermal variation of the glass volume — an elevation of temperature will cause an increase in the population of the right wells having a higher energy and corresponding to a larger glass volume. The subsequent quenching of a sample will be followed by a relaxation process in which overbarrier transitions from the right to the left well are prevailing over reverse transitions. As a result, the specific volume decreases towards its equilibrium value.

Notice that the positive sign of the parameter ***b***, established by us for polystyrene, does not mean that in any other glass ***b*** is also positive. Strictly speaking, it cannot be absolutely excluded that in some glasses ***b*** may have a negative value. In this connection we would like to remind that vitreous silica exhibits a *negative* thermal expansion at low temperatures ($T$ < 200 K). As known, at these temperatures TLSs give a remarkable contribution to the properties of this glass. Therefore one cannot rule out the possibility that this sign of the volume change with temperature in vitreous silica is due to TLSs having negative ***b***. If so, then the number of nearly symmetric TLSs, according to our theory, should increase with increasing pressure at small *P* values (unlike the decrease of the number of such TLSs in our system, glassy polystyrene). This conclusion is in agreement with an analogous conclusion made in Ref. 45, where under pressure an enhancement of the Brillouin scattering in vitreous silica at relatively low pressures up to about 20 kbar was observed.

Recently, an interesting study was published[46] on the pressure dependence (up to 28 kbar) of the photon echo in a polymer glass — polyisobutylene doped with tetra-tert-butylterrylene. The authors observed shortening of the photon echo decay time with increasing pressure, which is equivalent to increase in the pure homogeneous dephasing rate for radiative electronic transitions in an impurity molecule, or equivalent to *increase* in the homogeneous width $\Gamma_{hom}$ of the respective zero-phonon line. At first glance, this result seems to be inconsistent with the pressure-induced *decrease* of $\Gamma_{hom}$ as follows from our SHB studies of chlorin-doped glassy polystyrene[9,11-13] using the well-known relation $\Gamma_{hole} \approx 2\Gamma_{hom}$. However, one should keep in mind that the mentioned experiments were performed under very different conditions: the photon echo was studied at pressures from 0 to 28 kbar on a time scale of a few nanoseconds over the temperature range from 1.2 to 2.8 K (see Sec. III D of Ref. 46), where the tunneling-governed dynamics is expected to dominate, whereas our SHB experiments were made on a time scale of about $10-1000$ seconds and at higher temperatures, namely at $4.2-15$ K in Refs. 11-13 and at $5-18$ K in the present paper, when thermally activated processes increasingly contribute (not to mention dissimilar polymer glasses studied). Probably the most essential is the difference in pressures — in Refs. 9,11-13 they belong to the range of linear dependence on $P$, while in Ref. 46 they are clearly out of this range. Therefore, to interpret the results of photon echo experiments it is necessary to include into consideration also higher order terms in $P$.

## V. SUMMARY AND CONCLUSIONS

In brief, we have measured irreversible excess broadening of spectral holes in chlorin-doped polystyrene glass as a result of temperature cycling from 5 K to various temperatures up to 18 K and back to 5 K at different pressures up to 5 kbar. The experimental findings are: (i) a weakly superlinear ($\propto T^{3/2}$) dependence of the hole width increment on the cycling temperature, and (ii) a noticeable decrease of this increment with increasing pressure. These results fit well into a developed theoretical model where the hole broadening is assumed to be caused by irreversible spectral diffusion arising from coupling of the electronic transition in the impurity molecule to two-level systems in which thermally activated overbarrier jumps from one potential well to another take place.

The theoretical treatment of the observed effects is given within the framework of the soft anharmonic potential (SAP) model assuming (i) linear interaction of TLSs with the elastic strain caused by hydrostatic compression and (ii) asymmetric distribution of the cubic anharmonicity parameter *x* with respect to the sign of *x*. We have shown that our observations reflect a fundamental property of condensed matter — dissymmetry between compression and expansion (or decrease and increase of the volume). In glasses this dissymmetry can be taken into account within the SAP model if one accepts that there exists a correlation between the direction of change (or the sign) of the coordinate $x$ and the direction of variation of the glass volume $V$; namely, if $x$ changes then $V$ changes as well. Due to this correlation and the above dissymmetry, the distribution function for the cubic anharmonicity parameter $x$ turns out to be asymmetric.

It was shown that the observed by us pressure effect in glassy polystyrene can be explained if one accepts that the product of $x$ and of the harmonic oscillator force constant $h$ has for the majority of TLSs the negative sign, i.e. $hx < 0$. This means that single-well potentials of the soft localized modes (SLMs), existing only at positive values of $h$, must have $x_m < 0$, i.e. the most abundant are SLMs with *negative* $x$ values.[11] By contrast, double-well potentials, belonging to TLSs with negative $h$, must have $x_m > 0$ and hence also $\Delta_m > 0$, where $x_m$ is the most probable value of $x$ and $\Delta_m$ is the most probable asymmetry energy for TLSs. In other words, we have shown that *the most abundant low-energy TLSs in glassy polystyrene have double-well potentials with positive $x$ values and thus also with positive $\Delta$ values, i.e. the most numerous are double-well potentials for which the well corresponding to a larger volume of glass is sited at a higher energy than the another well*. External pressure increases the algebraic value of $\Delta$, reducing in this way the number of TLSs with small $|\Delta|$. Reduction in the number of such TLSs is the cause of the observed effect — a decrease of the residual broadening of spectral holes in temperature cycling experiments due to application of hydrostatic pressure.

As discussed above, our results are also consistent with other experimental facts, and such a treatment may be useful for interpreting further experimental observations performed at moderately high pressures in different time scales and temperature ranges.

## ACKNOWLEDGMENTS

The authors gratefully acknowledge financial support from the Estonian Science Foundation (Grants No. 3873, 4208, 5023, 5544) as well as from Kami Foundation (Sweden) and the U.S. National Research Council Twinning Program with Estonia.